\documentclass[twocolumn,bibnotes,aps,prl,floatfix]{revtex4-2}

\usepackage[english]{babel}
\usepackage[utf8]{inputenc}
\usepackage[T1]{fontenc}

\usepackage{graphicx}
\usepackage{bm}
\usepackage{comment}

\usepackage{physics}
\usepackage{amsmath}
\usepackage{float}
\usepackage[pdftex]{hyperref}
\usepackage{xcolor}
\usepackage{lineno}
\newcommand{\red}[1]{{\color{black}#1}}
\begin{document}


\title{Genuine Continuous Quantumness}

\author{Vojt\v{e}ch Kala$^\dagger$}
\email{kala@optics.upol.cz}

\author{Ji\v{r}\'i Fadrn\'y}
\thanks{These three authors contributed equally}
\author{Michal Neset}
\thanks{These three authors contributed equally}
\author{Jan B\'ilek}
\author{Petr Marek}
\author{Miroslav Je\v{z}ek}\email{jezek@optics.upol.cz}
\affiliation{%
 Department of Optics, Palack\'y University, 17. listopadu 1192/12, 77900 Olomouc, Czech Republic}

\date{\today}

\begin{abstract}
\section{Abstract}
Randomness is a key feature of quantum physics. Heisenberg's uncertainty principle reveals the existence of an intrinsic noise, usually explored through Gaussian squeezed states. Due to their insufficiency for quantum advantage, the focus is currently shifting towards genuinely quantum non-Gaussian states. However, while genuine quantum behavior comes naturally to discrete variable systems, its preparation and verification are difficult in continuous ones. Simultaneously, a unifying theoretical framework based on the continuous nature is missing. Here, we introduce nonlinear squeezing as a general framework to describe and verify genuine quantumness in the noise of continuous quantum states. Using this approach, we certify the non-Gaussianity of experimentally prepared multi-photon-added coherent states of light for the first time. Chiefly, we demonstrated the nonlinear squeezing corresponding to third- and fifth-order quantum nonlinearities, going significantly beyond the current state-of-the-art in quantum technology. \red{This framework enables uncovering intricate quantum properties in cutting-edge experiments and provides an efficient tool for further development of quantum technologies.}
\end{abstract}
\maketitle
\def\thefootnote{\dagger}\footnotetext{These authors contributed equally to this work}\def\thefootnote{\arabic{footnote}}

\section{Introduction}

While classical determinism allows complete predictions
about a physical system from the knowledge of its position and momentum, quantum physics does not even allow their simultaneous definition, much less measurement and utilization.
The intrinsic noise lies at the core of many effects \cite{Wilson2011,Hakonen,Almheiri,Marti2024,Noda2007,Nova2019,Crispino2008} and its inevitable presence impose fundamental limitations \cite{Giovannetti2004,Mason2019,Caves,Ghne2023,Menicucci}. It is a driving force of many rudimentary applications \cite{Maiman1960,Hepp1973,Kaminski2002}. Moreover, the superposition of many possibilities that together constitute the quantum noise can be utilized for the most complex of quantum physics applications \cite{Lloyd1999,Gottesman2001,Jia2024,Menicucci}. Those require and explore genuinely quantum states \cite{Kudra2021,Konno2024,Konno_nlsq,Eriksson2024,Fadrny2024,Nagata2007,Ogawa2016}. 

While genuine quantum behavior comes naturally to discrete variable systems, the vast palette of continuous genuinely quantum states lacks a unifying and suitable framework that would allow for their description and verification. Yet, the ability to identify truly quantum features in experimentally prepared states is of the same importance as the preparation itself \cite{OBrien2007,Varnava}. The infinite complexity of the continuous space of quantum states and ubiquitous imperfections that burden experimental data \cite{WalschaersPRXQ,Frattini2017,Kudra2021,Kala22,Konno2024} makes the task even more challenging.

Historically, such states were recognized by the presence of negativity in quasiprobability distributions describing the intrinsic noise \cite{LvovskyPRL,Ogawa2016,LvovskyTomo}, but this is only a binary condition that does not provide detailed information about the nature of the state. A more quantitative approach draws on the discrete number of energy quanta \cite{Straka2018,Chabaud2020,Lachman2022,Fiurek2022}, completely disregarding the continuous aspect.

In this paper, we introduce generalized nonlinear squeezing as a universal framework that enables to explore various shapes of noise distribution and verify their genuine quantumness. The concept has a direct visual interpretation via cost functions that penalize any deviation of the noise distribution from a target shape and unveils genuine quantumness in continuous systems without resorting to counting energy quanta. \red{As such, the nonlinear squeezing can be used to quantify excess noise in a realization of some ideal target state class and thus its quality and usefulness.} 

We show its applicability to quantum states prepared with current experimental technology. Only very recently, 20 years after the first preparation of single-photon-added coherent states \cite{Zavatta2004}, their generalization to $n$-photon-added coherent states was experimentally realized \cite{Fadrny2024}. The states are of fundamental \cite{Zavatta2004} and quantum informational importance \cite{Neset2024}, however, the certification of their quantum non-Gaussianity remained unresolved.

We experimentally prepare single- and two-photon added-coherent states and certify their quantumness via first time observation of the nonlinear squeezing in multiphoton states. Specifically, we explore the cubic nonlinear squeezing that produces a class of states vital for photonic quantum computing \cite{Lloyd1999,Budinger}. Furthermore, we use the introduced concept of generalized nonlinear squeezing to define a quintic nonlinear squeezing and report the first observation of nonlinearity of higher than cubic order in the quantum regime. 

\section{Results}
\subsection{Continuous quantum noise}

In classical physics, systems can be described in phase space by simultaneously tracking the exact values of their positions and momenta. This is not allowed in quantum physics, where the position and momentum are represented by non-commuting operators, and their precise values cannot be simultaneously defined nor measured. The theory has a peculiar consequence---the ground state of an oscillator with zero mean position and momentum allows observation of their nonzero values. To give an example, for the vacuum state of light, in the absence of photons, nonzero values of the electromagnetic field can be measured.

The uncertainty in position and momentum renders inherent quantum noise that can be described by a quasiprobability distribution in conjugated variables, such as the Wigner function. For the ground state of a harmonic oscillator, the Wigner function gains the form of a two-dimensional Gaussian distribution located in the origin of the phase space. The Heisenberg uncertainty relation limits the product of its variances. However, it does not forbid suppressing the noise carried by one variable while the variance of the conjugated variable increases. Such suppression can be found in the Gaussian squeezed states \cite{Andersen2016}. Since their first experimental preparation in 1985, squeezed states have found many applications \cite{Aasi2013,Zhong1460,Asavant,Larsen}. 

\red{Despite their ability to reach noise levels lower than any classical state, Gaussian squeezed states do not possess two key quantum properties. First, Gaussian states can be efficiently simulated on classical computer \cite{Mari2012}, meaning that a quantum computer consisting solely of them and linear operations preserving their Gaussianity does not provide any computational advantage based on quantum physics. Second, their Wigner function is positive, absent of interesting negative regions, and enabling formulation of local hidden variable models for Gaussian measurements \cite{Jabbour2023}.} 

To obtain a system that resists \red{efficient} classical simulation and thus can possess a quantum computational advantage, it is necessary to go beyond the set of Gaussian squeezed states into the genuinely quantum regime. We define genuine quantum states as states with non-Gaussian noise distribution that cannot be written as a classical statistical mixture of Gaussian states \cite{WalschaersPRXQ}. For illustration, the definition draws a line between a bimodal non-Gaussian Wigner function of a bistable laser and a quantum non-Gaussian superposition of two coherent states named optical Schr{\"o}dinger cat state \cite{Ourjoumtsev2006,Ourjoumtsev2007}. 

As opposed to a Gaussian state that enables a parametrization via mean values and a covariance matrix, a generic quantum non-Gaussian state cannot be \red{fully} described by a finite number of parameters. To define a framework capable of grasping the complex space of continuous quantum non-Gaussian states, we define generalized nonlinear squeezing.

\subsection{Generalized nonlinear squeezing}

Let us first revisit the Gaussian quadrature squeezing
in a single-mode harmonic oscillator described by
position $x$ and momentum $p$ quadrature operators with
$[x, p] = i$ \red{$(\hbar = 1)$}. Gaussian squeezing is commonly evaluated by a variance of one of the conjugate variables, $x$, for example. For quantum states with zero mean value of $x$, the variance comes down to the second moment. It is sufficient to discuss this case, because $\expval{x}=0$ can be always set by a suitable displacement operation that does not change the shape of noise distribution. The variance is then compared to its minimum value over the set of classical states consisting of coherent states and their mixtures, which is simply the variance in the vacuum state.

\begin{figure}
  \includegraphics[width=0.8\columnwidth]{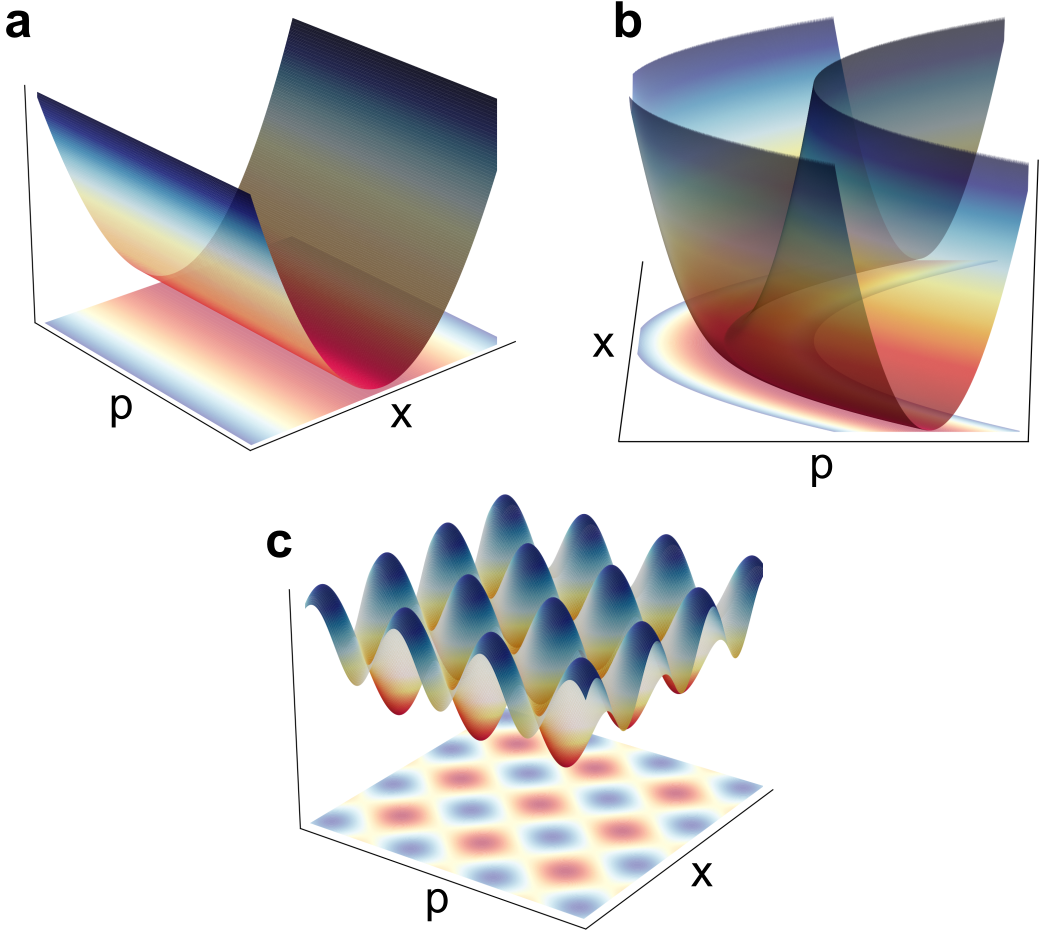}
  \caption{\textbf{a}, Cost function $f(x,p)^2 = x^2$ for the Gaussian squeezing. It forms a narrow valley that follows a straight line. \textbf{b}, Cost function for the nonlinear cubic squeezing with $f(x,p)^2=(p+zx^2)^2$. Its valley follows a parabola curve. \textbf{c}, Cost function for the nonlinear GKP squeezing. Its minima lies on a grid and the function resembles an egg carton.}
  \label{costfunctions}
\end{figure}

In terms of Wigner function $W(x,p)$ describing the quantum system $\rho$, the second moment of position equals
\begin{equation}\label{varGauss}
    \expval{x^2} = \int\!\textrm{d}x \textrm{d}p\, x^2 W(x,p).
\end{equation}
The function $x^2$ forms a narrow valley that follows the $p$ axis in phase space, see Fig.~\ref{costfunctions}\textbf{a}. The overall integral is small when the larger values of the Wigner function are concentrated along the bottom of the valley. When seen as a cost function, the valley penalizes any volume of the distribution spreading in the $x$ direction. The variance \eqref{varGauss} shows the presence of the Gaussian squeezing when smaller than the variance in the vacuum state.

The Gaussian squeezing can be generalized to nonlinear squeezing by allowing for more complex shapes of the cost function $f(x,p)^2$, given by an arbitrary nonlinear function $f(x,p)$. The corresponding variance reads
\begin{equation}\label{varf}
    \textrm{variance}_\rho\left(f(x,p)\right)= \expval{f(x,p)^2}-\expval{f(x,p)}^2.
\end{equation}
In analogy to Gaussian squeezing, where the variance is compared to the minimum over classical states, we define the nonlinear squeezing $\xi$ of the quantum state $\rho$ as a ratio of variance \eqref{varf} and its minimum over Gaussian states and their mixtures,
\begin{equation}\label{definition}
    \xi = \frac{\textrm{variance}_{\rho}(f(x,p))}{\min_G \textrm{variance}_{\rho_G}(f(x,p))}.
\end{equation}
Thus, any state $\rho$ with $\xi<1$ is quantum non-Gaussian. In contrast to the quantum non-Gaussianity based on counting energy quanta \cite{Straka2018,LachmanG19,Chabaud2020,Lachman2022}, nonlinear squeezing reveals the genuine quantumness in the noise of continuous variables.

The Wigner function of the given state can be arbitrarily placed in the phase space. Still, the positioning does not affect the Gaussian or quantum non-Gaussian shape of the distribution. To focus solely on the shape of the Wigner function and find the minimum attainable nonlinear squeezing with a given state, we should appropriately rescale the cost function. This is equivalent to applying a generic Gaussian unitary operation $U_G$ to the state under consideration,
\begin{equation}\label{UG}
    \rho \rightarrow U_G\rho U_G^{\dagger}.
\end{equation}
The transformation only linearly changes the variables of the Wigner function. The minimum nonlinear squeezing for the state $\rho$ is then revealed via an optimization of the operation $U_G$, see Methods for more details.
Independently of the chosen nonlinear function $f(x,p)$, the nonlinear squeezing forms a witness of non-Gaussianity when optimized over the transformation $U_G$. Simultaneously, this approach defines a structure in the space of the non-Gaussian states. Different cost functions generate classes of states whose Wigner functions follow the shape given by $f(x,p)$.
\red{Moreover, in direct analogy to the linear squeezing, the value of the nonlinear squeezing $\xi < 1$ quantifies the non-Gaussian resource within the quantum state that can be utilized in specific applications.}

The cost function connected to the Gaussian squeezing prioritizes states squeezed along a line. The next higher order curve in the phase space is parabola, with
\begin{equation}\label{f3}
    f(x,p) = p + z x^2.
\end{equation}
The real parameter $z$ is half of the inverse of the \mbox{semi latus rectum} of the parabola. The corresponding cost function is shown in Fig.~\ref{costfunctions}\textbf{b}. The ideal state squeezed according to \eqref{f3} is the cubic state that can be generated by the cubic Hamiltonian $H_3=\frac{\chi}{3}x^3$ \cite{Gottesman2001}. The Hamiltonian transforms the momentum $p$ as $p \rightarrow p + \chi x^2$.
The cubic nonlinearity and the corresponding nonlinear squeezing recently received a lot of attention.
The cubic state is a highly sought quantum state due to its ability to unlock universality within continuous variable quantum computing \cite{Sakaguchi2023,Eriksson2024}.
\red{Specifically, the cubic nonlinear squeezing is a vital non-Gaussian resource required in the implementation of the measurement induced cubic phase gate. The degree of the cubic squeezing of the ancillary state determines the amount of noise added during the process \cite{Miyata2016,Konno_nlsq,Kala22,Brauer2021,Eriksson2024}.}

Following this logic, we can define cost functions following more complex curves. To give an example the cost function for nonlinear quintic squeezing can be defined as
\begin{equation}\label{f5}
    f(x,p) = p + s x^2 + r x^4.
\end{equation}
Similar to the cubic squeezing, the chosen form is invariant under reflection $x \rightarrow -x$. It is possible to include also the term $x^3$ when considering states with different symmetry. Generally, including higher-order terms enables to refine the structure of the cost function and target states generated by higher nonlinear processes and complex measurement-induced preparations. It is not necessary to limit ourselves to cost functions that follow a one-dimensional curve in the phase space. Recently introduced nonlinear squeezing for Gottesman-Kitaev-Preskill states \cite{Marek2024} can be interpreted as using a cost function whose minima lie on a periodic grid resembling an egg carton, see Fig.~\ref{costfunctions}\textbf{c}. \red{Similar characterization was developed for Schr\"odinger cat states \cite{Catability,Kuchar2025}. In both cases, the ground state nature of the studied state with respect to a specific operator was utilized to define the nonlinear squeezing. For more notes on the choice of the cost function, see Methods.}

\subsection*{Experiment: photon-added coherent states}

We demonstrate experimentally for the first time the nonlinear squeezing of multiphoton quantum states. We study photon-added coherent states, which are the result of repeated action of the creation operator $\hat{a}^\dagger$ on a coherent state $|\alpha\rangle$,
\begin{equation}\label{PACSdefinition}
|\alpha,n\rangle=\mathcal{N}_n(\alpha) \hat{a}^{\dagger n} |\alpha\rangle,
\end{equation}
where $\mathcal{N}_n(\alpha)=[\langle \alpha| \hat{a}^n\hat{a}^{\dagger n}|\alpha\rangle]^{-1/2}$ is the normalization factor and $n$ stands for the number of added photons. These states are non-Gaussian and feature complex interference in the phase space with positive and negative areas of their Wigner functions.

The peculiar property of the photon-added coherent states lies in the composition of their energy. The added nonclassical energy quanta is accompanied by the strong classical drive, and the difficulty of showing the quantumness of those states increases with the strength of the coherent component \cite{Zavatta2004}. On top of that, quantum states with a larger mean photon number are generally more susceptible to losses and other imperfections, which imposes extreme requirements on experimental precision \cite{Harder}.

\begin{figure}
\includegraphics[width=0.9\columnwidth]{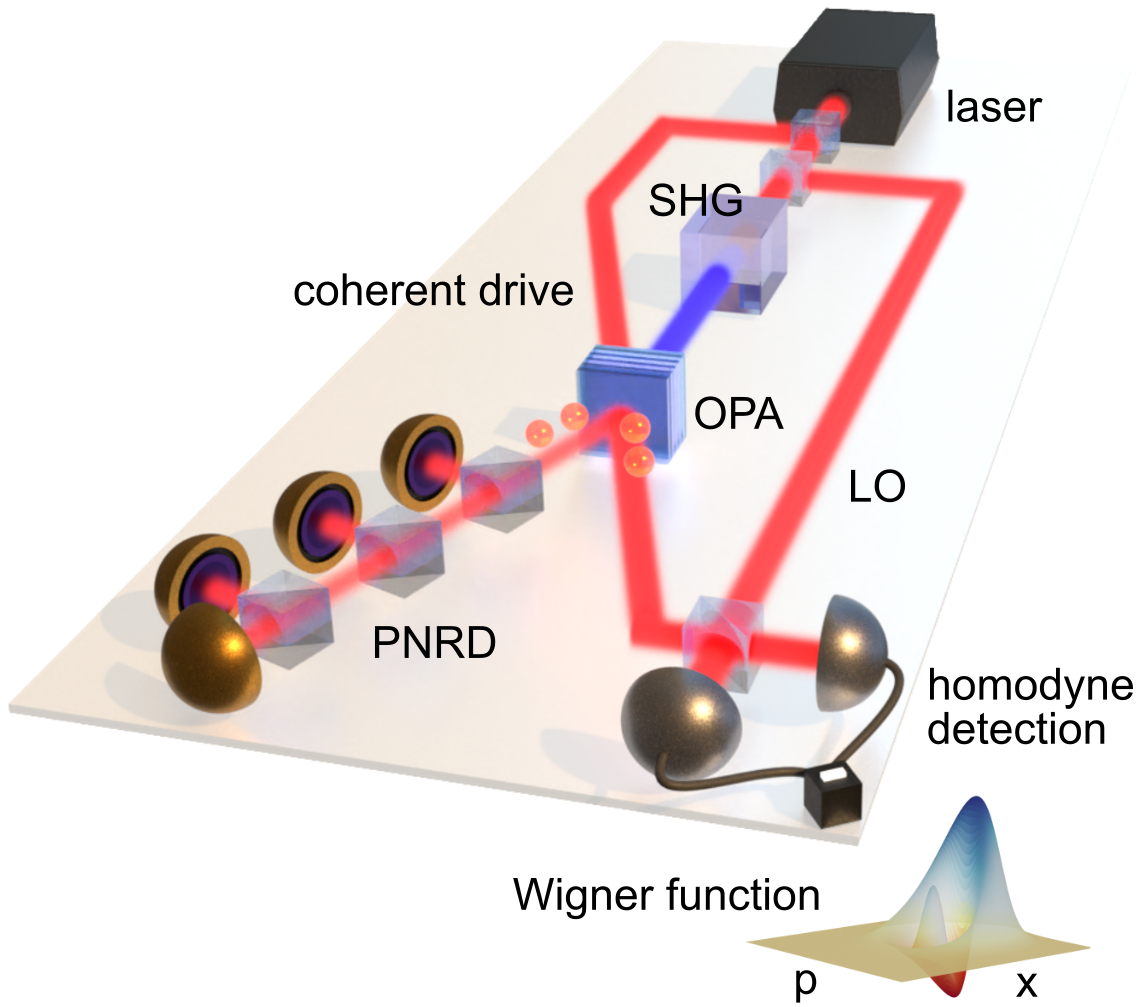}
\caption{Schematic of the  preparation of $n$-photon-added coherent states. A coherent state $|\alpha\rangle$ from a fundamental laser is seeded into the signal mode of an optical parametric amplifier (OPA). The OPA is pumped using a second-harmonic-generated light (SHG).  Detection of $n$ photons at the photon-number resolving detector (PNRD) in the auxiliary mode projects the signal mode to the desired $n$-photon-added coherent state. The resulting state is completely characterized by homodyne detection using a local oscillator (LO) and a balanced pair of photodetectors.}
\label{fig:experiment}
\end{figure}

\begin{figure}
\includegraphics[width=0.9\columnwidth]{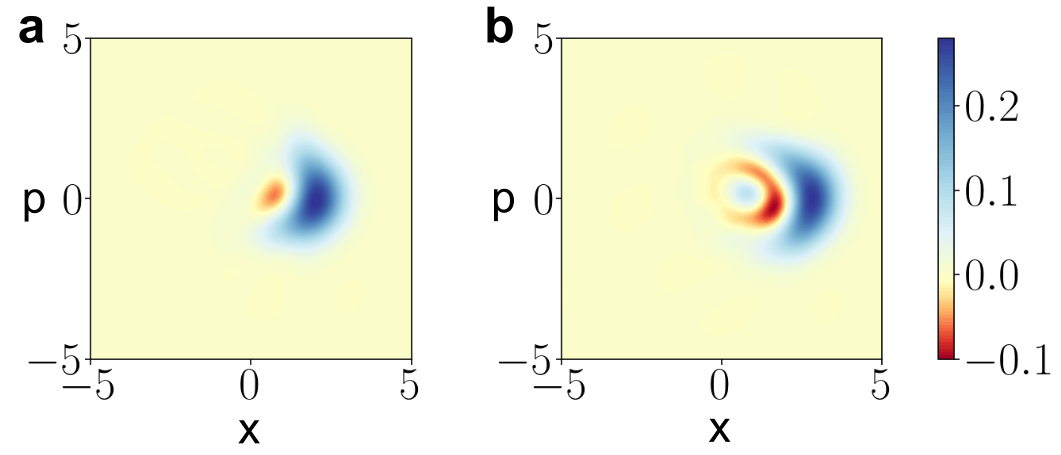}
\caption{Density plots of the Wigner function of the experimentally generated states: \textbf{a}, single-photon-added coherent state; \textbf{b}, two-photon-added coherent state. Both have the coherent drive with an amplitude $|\alpha|$ of approximately 1.0.}
\label{fig:measured_wigners}
\end{figure}

We experimentally prepared the multiphoton-added coherent states in a pulsed optical parametric amplifier (OPA) seeded by a coherent state, see Fig.~\ref{fig:experiment}. The amplifier consists of a nonlinear crystal pumped by a strong ultraviolet picosecond light produced in the process of second-harmonic generation. Upon the absorption of the pump photon, pairs of correlated red photons (signal and auxiliary) are emitted. The number of photons in the auxiliary mode is measured using a photon-number-resolving detector (PNRD) based on spatial multiplexing and single-photon detectors. Detection of $n$ photons in the auxiliary mode heralds the preparation of the same number of photons in the signal mode. The weak coherent state, derived from the fundamental laser, is spatiotemporally matched and injected into the input of the signal mode of the OPA. Consequently, the multiphoton addition is performed on the coherent drive. We fully characterized the prepared states by means of time-domain balanced homodyne detection and quantum state reconstruction. The examples of the measured Wigner functions of single- and two-photon-added coherent states are shown in Fig.~\ref{fig:measured_wigners}. The first experimental preparation of the two-photon-added coherent states reaches high fidelity of 97\% with the corresponding theoretical state \eqref{PACSdefinition}.

\begin{figure}[ht!]
\includegraphics[width=0.8\columnwidth]{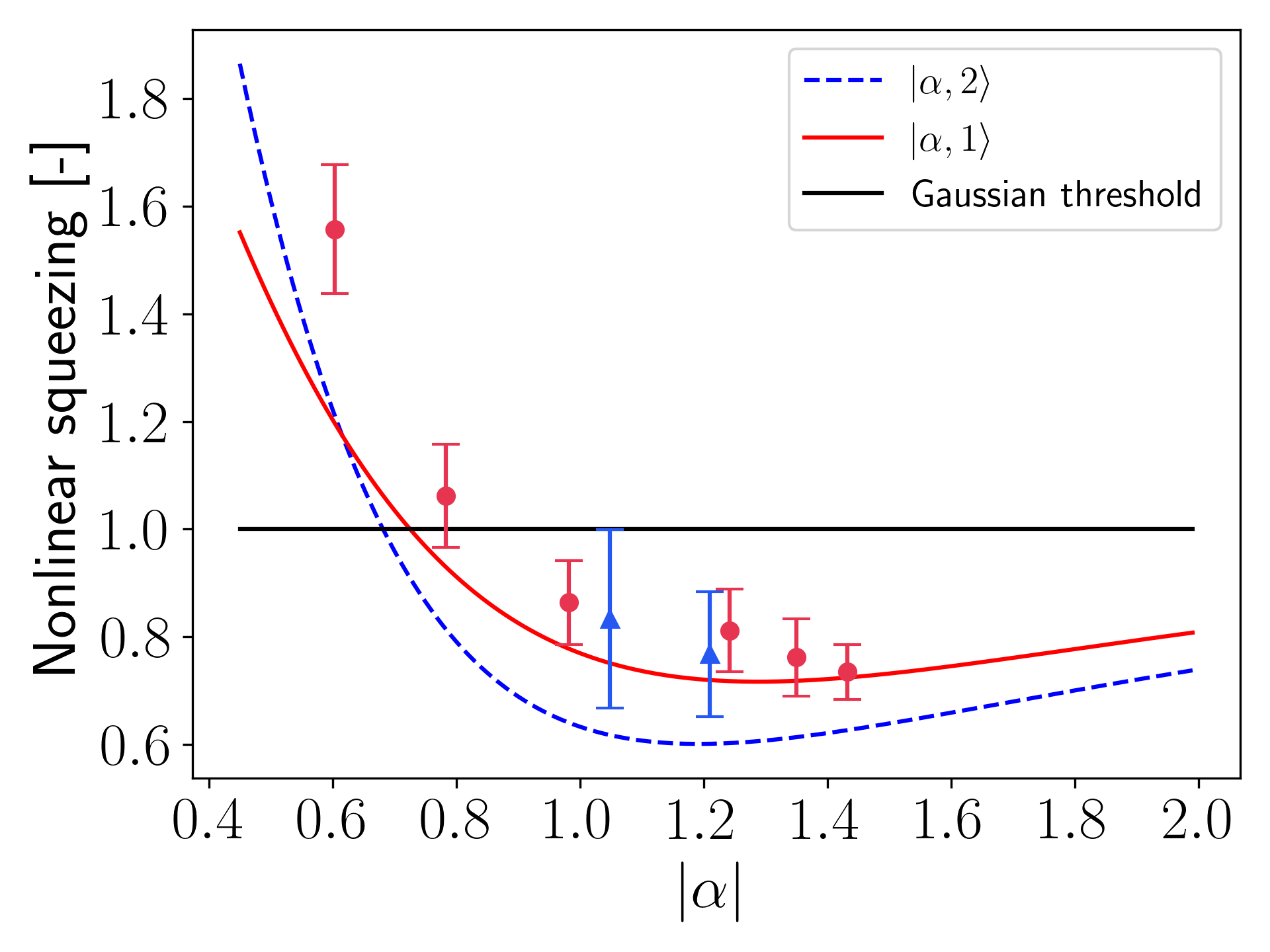}
\caption{Nonlinear cubic squeezing of photon-added coherent states. The theoretical prediction of cubic squeezing of single-photon-added coherent states, and two-photon-added coherent states are plotted as red solid, and blue dashed lines, respectively. Red dots, and blue triangles represent values of cubic squeezing experimentally determined from measured data for single-, and two-photon-added coherent states, respectively. The area below the black line contains the states with the nonlinear cubic squeezing. The error bars represent one standard deviation.}
\label{nls_results}
\end{figure}

We certify the quantum non-Gaussianity of the prepared state by witnessing the nonlinear squeezing. We evaluated the nonlinear squeezing \eqref{definition} with the cost function $f(x,p)^2$ given by \eqref{f3}, which corresponds to the cubic nonlinear squeezing. The variance of the measured state was optimized over all Gaussian operations \eqref{UG}-\eqref{gaussians}. We performed this certification for various numbers of added photons and the strength of the amplitude of the coherent state, see Fig.~\ref{nls_results}. Results depicted in red correspond to the single photon-added coherent states, and blue color describes two-photon-added coherent states. The markers show the nonlinear squeezing of experimentally prepared states and can be compared to solid lines showing the nonlinear squeezing of the ideal photon-added coherent states \eqref{PACSdefinition}. The black line represents a bound; values below this bound certify quantum non-Gaussianity. We successfully certified the quantum non-Gaussianity of four single-photon and two two-photon-added coherent states. Increasing the number of photon additions poses more stringent requirements on the experimental technology. We also prepared three-photon-added coherent states. Despite their high fidelity of 0.86, the certification of their non-Gaussianity via the cubic nonlinear squeezing was not possible, with the minimal value $\xi = 1.5 \pm 0.3$. 

These results demonstrate an unprecedented quality of coherent multiphoton addition reached despite many technological challenges. Achieving precise mode matching and, consequently, high purity of states is experimentally demanding, especially given that gain-induced diffraction and dispersion in the pulsed regime degrade the quality of the n-photon states. Optimizing pulse duration and focusing in the OPA requires a delicate balance: shorter pulses and stronger focus intensify undesirable effects, while longer pulses and weaker focus reduce the generation rate.
The high peak density of pump power can damage the nonlinear crystal, necessitating continuous shifting to maintain crystal integrity during higher-photon addition experiments. Furthermore, the picosecond operational regime requires advanced sub-nanometer filtering of the auxiliary mode achieved by holographic volume gratings. Finally, ensuring reliable state measurement demands an ultra-stable homodyne detector and precise synchronization with a high-throughput coincidence logic. See Methods for more details on the experiment.

\begin{figure}[ht!]
\includegraphics[width=0.9\columnwidth]{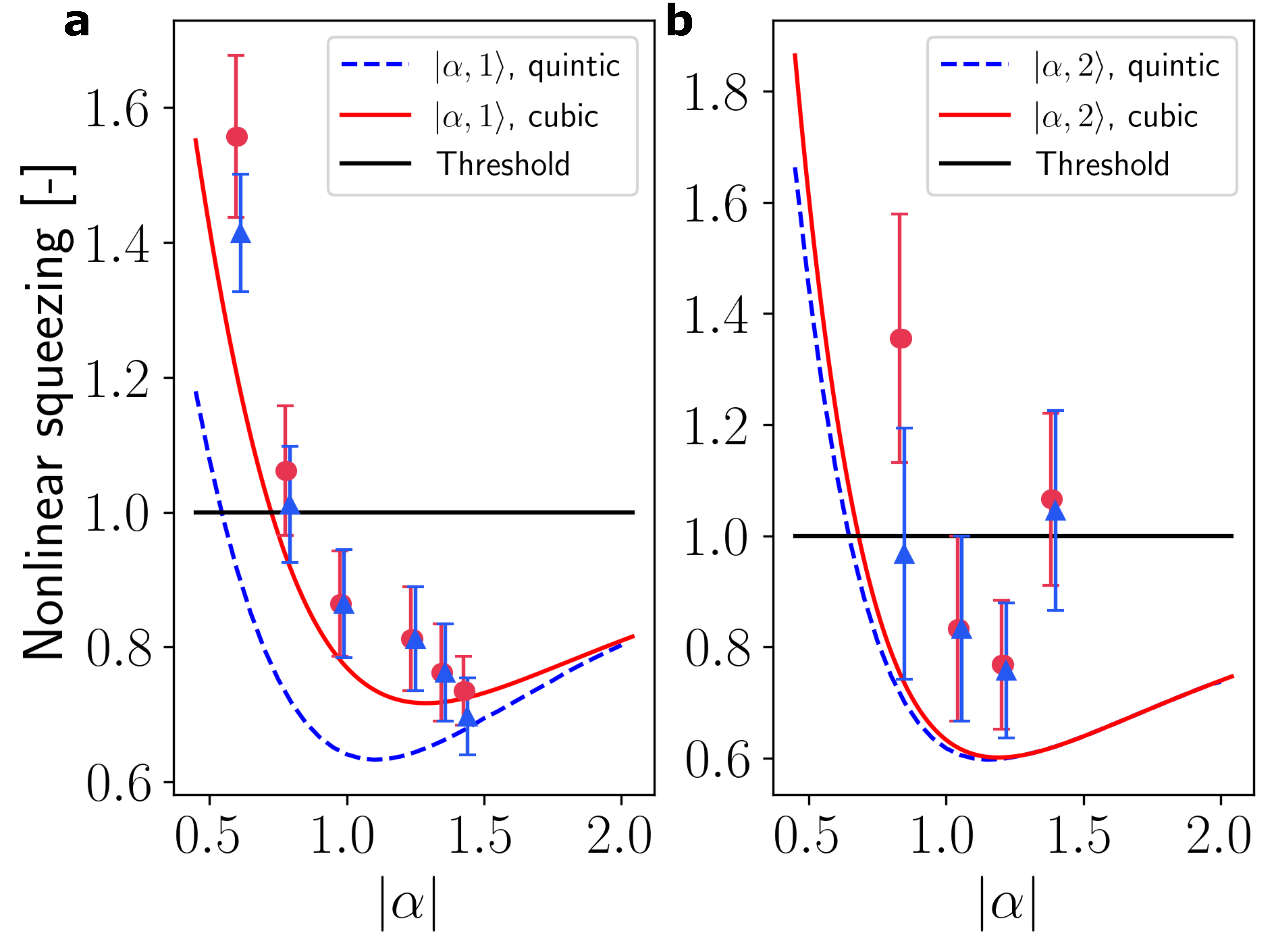}
\caption{Comparison of cubic (red solid lines) and quintic (blue dashed lines) nonlinear squeezing of single-photon (\textbf{a}) and two-photon (\textbf{b}) added coherent states. Red dots, and blue triangles represent values of cubic, and quintic squeezing, respectively, experimentally determined from measured states. All states manifest equal or higher quintic nonlinear squeezing compared to the cubic squeezing due to better matching between the cost function and distribution of quantum noise in measured states.}
\label{quintic}
\end{figure}

Certified two-photon-added coherent states seem like the limit of the experiment. Further improvement given the current experimental technology can be achieved by refining the cost function. In a similar manner to evaluation of the cubic squeezing, we analyzed the quintic squeezing \eqref{f5} for the experimentally obtained states.
\red{The quintic cost function expands the space of shapes that the quantum noise of the state may attain and be detected by the nonlinear squeezing, which increases the sensitivity of the method. The improvement can be seen in Fig.~\ref{quintic}.}
\red{Even though the theoretical prediction for single- and two-photon-added coherent states are rather similar, the experimental states are burdened by imperfections. A more complex cost function is thus more sensitive to the non-Gaussianity within their quantum noise. Specifically, for the two-photon-added coherent state with amplitude $\alpha = 0.84$, refining the cost function from cubic to quintic yields an improvement in the evaluated value of nonlinear squeezing from $1.36 \pm 0.22$ to $0.97 \pm 0.23$. In other words, the probability of detecting the genuine non-Gaussianity of this particular state is consequently increased from $22.6\%$ to $52.6\%$. Furthermore, the determined value of the quintic squeezing for the single-photon-added coherent state with amplitude $\alpha = 1.43$ goes beyond the sensitivity of the cubic squeezing alone.}

\section*{Discussion}

We presented a concept of general nonlinear squeezing for continuous quantum systems. It provides a framework for the description of various shapes of the inherent quantum noise. Utilizing the nonlinear squeezing, we defined a witness of genuine quantumness, or quantum non-Gaussianity, and used it for the first-time certification of the elusive non-Gaussianity in experimentally prepared multi-photon-added coherent states. We showed that the experimentally prepared states manifest cubic and quintic nonlinear squeezing, where the latter has never been observed before.

Furthermore, successfully certifying genuine quantumness in states prepared by photon addition demonstrates the fully quantum nature of the operation, enabling its use in future quantum information protocols, such as noiseless amplification \cite{Zavatta2011,Fadrny2024,Neset2024}. Despite the challenging requirements on precision both in experimental preparation of the state and its measurement, successful identification of quantum non-Gaussianity shows a working interplay between the nonlinear squeezing and state-of-the-art quantum experiments. Identification of new cost functions suitable for detecting genuine quantumness in other problems in quantum physics remains an important open question for further research.

\section*{Methods}

\textbf{Theory:}
Choice of the cost function: two basic properties are required from the cost function a) existence of the mean values in \eqref{varf}, which requires existence of corresponding integrals and b) positivity of the minimum obtained over the set of Gaussian states.

\red{Formulating a single necessary and sufficient criterion for non-Gaussianity in mixed states is an extremely challenging task. Therefore, the nonlinear squeezing works in a similar way to entanglement witnesses, providing a sufficient condition for a class of non-Gaussian states. Thanks to the generalization presented here, it is possible to define different cost functions and address different classes of non-Gaussian states. However, the process of choosing a suitable cost function can benefit from prior information. Fortunately, this is often the case in quantum science and technology, where we either target a specific state (with application potential) with an experimental scheme or we deal with a known experimental processes, that yields a known outcome, yet with quantum properties diminished by imperfections and decoherence.

The prior information can be used in definition of the cost function. If the ideal studied state is an eigenstate of some operator, it yields a zero variance with respect to this operator, which can be used for definition of the cost function. The nonlinear squeezing of the experimentally prepared approximation can then be interpreted as a quality with which it approximates the target state.

In the other case, when the operator is not known, or it has a zero variance in some Gaussian state, we can explore symmetries possessed by the studied state. We used this approach here when defining the quintic squeezing for characterization of the photon-added coherent states.
}

The nonlinear squeezing is evaluated for an experimentally obtained state $\rho$ with a Wigner function $W(\textbf{r})$, 
\red{
\begin{equation}\label{nlsqWig}
\begin{split}
    &\xi(z) =\\& \frac{\min_{M,\textbf{d}}[\int W(M\textbf{r}+\textbf{d})f(\textbf{r})^2 \textrm{d}\textbf{r}-(\int W(M\textbf{r}+\textbf{d})f(\textbf{r}) \textrm{d}\textbf{r})^2]}{\min_{G}[\int W_G(\textbf{r})f(\textbf{r})^2 \textrm{d}\textbf{r}-(\int W_G(\textbf{r})f(\textbf{r}) \textrm{d}\textbf{r})^2]},
    \end{split}
\end{equation}
where $\textbf{r}=(x,p)$, matrix $M$ corresponds to  symplectic transformations of the phase space and \textbf{d} to displacements.} Wigner function $W_G$ is a Wigner function of a quantum Gaussian state. 
Alternatively to Wigner picture, Weyl transform $\hat{O}$ of \eqref{f3} and \eqref{f5} can be used for the cases of cubic and quintic nonlinear squeezing, yielding 
\begin{equation}\label{nlsqTr}
    \xi(z) = \frac{\min_{U_G}[\textrm{Tr}[U_G\rho U_G^{\dagger}\hat{O}^2]-\textrm{Tr}[U_G\rho U_G^{\dagger}\hat{O}]^2]}{\min_G[\textrm{Tr}[\rho_G\hat{O}^2]-\textrm{Tr}[\rho_G\hat{O}]^2]}.
\end{equation}
In the nominator, we perform the optimization over all Gaussian unitary operations $U_G$, where $U_G$ consists of rotation $R=\exp(i\phi a^{\dagger}a)$, displacement $D=\exp(\alpha a^{\dagger} - \alpha^{*}a)$, and Gaussian squeezing $S=\exp(\frac{r}{2}(a^2-a^{\dagger 2}))$,
\begin{equation}\label{gaussians}
    U_G = R(\theta)D(\alpha)S(r)R(\phi).
\end{equation}

\red{Generally, a full quantum state reconstruction is not necessary to estimate the nominator in \eqref{nlsqWig}-\eqref{nlsqTr}, if an active phase locking of the homodyne measurement is available. It is not the case of our experiment as it was performed in a pulsed regime, where the phase of the measured data was estimated, but not actively locked beforehand (see Methods: Experiment).

However, if the phase locking is available, the moments constituting the nonlinear squeezing can be estimated directly from homodyne measurements under specific angles as  \cite{Moore,Kala2025,NonGNull}. Specifically, Weyl symmetric moments can be composed of moments of rotated quadratures as \cite{NonGNull}
\begin{equation}\label{Wsym}
  :x^mp^n:_W  = \sum_{k=1}^{m+n} A_k X(\theta_k)^{m+n},
\end{equation}
where $X(\theta) = \cos(\theta)x + \sin(\theta)p$.}

\red{The denominator in \eqref{nlsqWig}-\eqref{nlsqTr} is minimized over the set of Gaussian states and their mixtures. However, it has been shown that a variance is minimized by pure states \cite{Kala22}, therefore, it suffices to consider pure Gaussian states, which can be parameterized by their first and second moments, the covariance matrix and mean values of $x$ and $p$. The denominator can then be expressed as a function of elements of the covariance matrix and the mean values and minimized. In the case of the cubic squeezing \eqref{f3} the minimum is found analytically in squeezed vacuum state yielding \cite{Kala22} 
\begin{equation}
\min_G\textrm{variance}_{\rho_G} = \frac{3}{2^{ \frac{5}{3}}}  |z|^{\frac{2}{3}}.
\end{equation}

In the case of the quintic squeezing \eqref{f5}, the threshold expressed in terms of covariance matrix and mean values is optimized numerically. Due to the form of the definition, the nonlinear quintic squeezing does not depend on displacement in the $p$ direction and is symmetric in $x$. Theoretical curve of the nonlinear quintic squeezing of the single- and two-photon-added coherent states is shown in Fig.~\ref{quintic}. 

Considering the free parameters in the nonlinear squeezing, the parameter $z$ changes the shape of the parabola curve followed by the cubic cost function. It should be chosen such that the nonlinear squeezing is minimal. Adjusting the parameter $z$ is equivalent to applying Gaussian squeezing on the state under consideration. The additional parameter in \eqref{f5} of quintic squeezing enables a more sensitive characterization of the photon-added coherent states. Both parameters $s$ and $t$ are optimized, such that the maximal nonlinear squeezing is revealed. }

\textbf{Experiment:}
The fundamental coherent light is derived from a pulsed Ti: Sapphire laser (Coherent Mira HP), outputting 1.5 ps pulses at a central wavelength of 800 nm with a repetition rate of 76 MHz and an average power of 3 W. The majority of the laser power is down-converted in the process of second-harmonic generation (SHG) to pump a single-pass optical parametric amplifier (OPA). The OPA is based on the collinear degenerate type-II second-order nonlinear interaction in a 2mm long periodically poled potassium titanyl phosphate (PPKTP) crystal (Raicol Crystals). Pairs of correlated photons produced in the OPA have orthogonal polarizations. Once the residual pump is filtered out by a cut-off filter, photon pairs are separated by a polarizing beam splitter to signal and auxiliary modes.
The heralding efficiency of the source determined from independent measurement is 98\% when corrected for technical losses. The fundamental laser is attenuated and injected into the OPA as a coherent seed. It matches the spatio-temporal mode of the signal. At the output of the OPA, the auxiliary mode undergoes spatial and 0.4nm spectral filtering using single-mode optical fiber and holographic volume grating. After the filtering, it is detected by a photon-number-resolving detector (PNDR) consisting of a balanced free-space multiplexed network of single-photon detectors. Detection of a particular number of photons heralds the successful generation of the corresponding photon-added coherent state. Complete homodyne tomography of generated states is performed by a time-domain balanced homodyne detector (BHD) with record performance in terms of balancing stability, keeping other specifications on par with the state-of-the-art. Notably, the quantum efficiency of our detector is 92\%, the signal-to-noise ratio is around 12 dB, and the bandwidth exceeds 100 MHz. The homodyne detector reaches a long-term balancing stability of several hours. The local oscillator (LO) for the homodyne detector is derived from the fundamental laser. The detected signal is digitized by a high-speed oscilloscope (Teledyne LeCroy) in a memory-segmentation regime and transferred to a computer. \red{A density matrix and Wigner function of the measured state of light is reconstructed by the maximum-likelihood estimation, including the correction for technical and detection losses, yielding the physically sound reconstructed state}\cite{Jezek2003,Lvovsky2004,Fadrny2024}.
\red{We have achieved the fidelity of the generated single- and two-photon-added-coherent states as high as 97\%. Even though it represents an unprecedented high quality of the generation of this class of quantum states, the prepared states suffer from a few experimental imperfections.
Certification or even quantification of non-Gaussianity becomes significantly more difficult when the state experiences losses and other experimental imperfections. Furthermore, the photon-added coherent states are progressively more difficult to be certified non-Gaussian for high amplitudes of the coherent state as the classical coherent component of the state becomes dominant \cite{Fadrny2024}.
The main source of imperfection in our experiment is a phase noise that causes the Wigner functions to be slightly unsymmetrical, which affects the achievable nonlinear squeezing. 
While these imperfection are not apparent from the fidelity as the figure of merit, the proposed framework of nonlinear squeezing precisely quantifies the amount of advanced non-Gaussian resource within the quantum state, which is linked to its applicability in a particular quantum information processing protocol.
At top of that, various cost functions can be analyzed to further enhance the sensitivity of the method.
The uncertainty of the estimation process is carefully analyzed by Monte Carlo simulations seeded by the experimentally measured quadrature samples. The error bars in all the figures show one standard deviation.}

\begin{figure}[ht!]
\includegraphics[width=0.9\columnwidth]{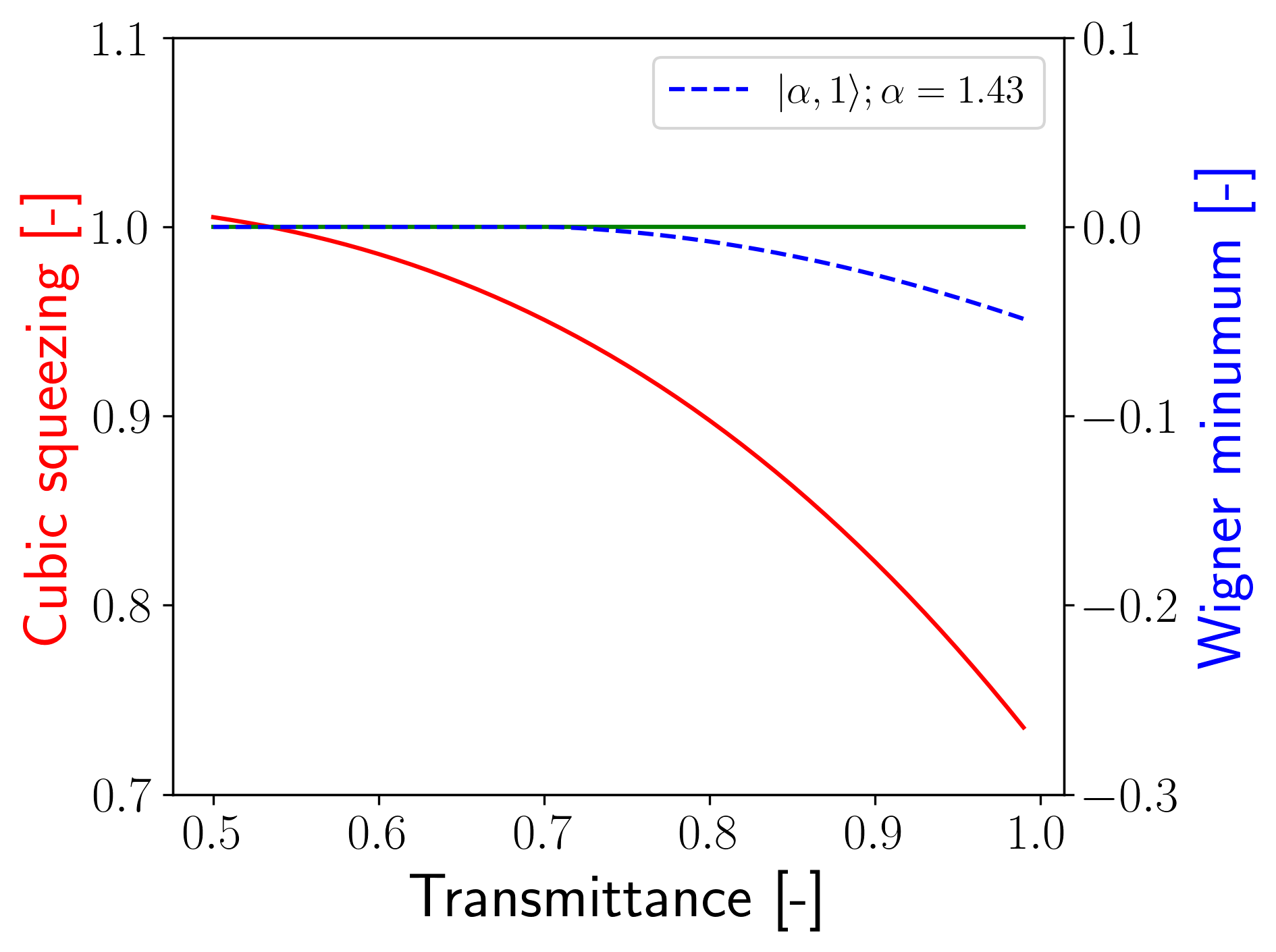}
\caption{\red{Dependence of the cubic nonlinear squeezing (red solid line) and Wigner function minimum (blue dashed line) of the single-photon-added coherent state with amplitude  $\alpha = 1.43$ on the optical losses in the system. While the negativity of the Wigner function has already vanished and can no longer certify non-Gaussianity, the state still features non-trivial cubic squeezing for transmittance as low as 0.6}}
\label{nlsq_vs_wig}
\end{figure}

\textbf{Comparison with the negativity of the Wigner function:}
\red{Negativity of the Wigner function clearly shows non-Gaussianity of pure quantum states. The number of disjoint negative regions in the Wigner function may indicate the degree of non-Gaussianity, but it does not generally identify the studied state. Once considering also statistical mixtures, completely different types of states can manifest the same number of negative regions. For instance, a Fock state and an approximate GKP state can have the same number of disjoint negative regions in their Wigner function while being of a completely different nature.
Furthermore, the negativity of the Wigner function is more susceptible to losses in comparison with nonlinear squeezing. Specifically, the negativity of the Wigner function of lossy photon-added coherent states vanishes quickly for higher aptitudes of the coherent state. Fig.~\ref{nlsq_vs_wig} visualizes the theoretical cubic squeezing and Wigner function minimum of a single-photon-added coherent state with amplitude 1.43. While the negativity of the Wigner function has already vanished, non-Gaussianity of the state can still be certified for transmittance as low as 0.6. The minimal level of transmittance required to certify non-Gaussianity of a single-photon-added coherent state further decreases for higher amplitudes of the coherent state.}



\section*{Data availability}
The data that support the findings of this study are publicly available on GitHub \cite{github}.

\section*{Code availability}
The codes used to process data and generate figures are publicly available on GitHub \cite{github}.

\section*{Acknowledgements}
JF, MN, JB, and MJ acknowledge the financial support of the Czech Science Foundation (project 21-23120S). JF, MN, and VK acknowledge project IGA-PrF-2025-010 of the Palacký University. 
PM and VK acknowledge the financial support of the Czech Science Foundation (project 25-17472S), and European Union’s HORIZON
Research and Innovation Actions under Grant Agreement no. 101080173 (CLUSTEC). VK acknowledges the Quantera project CLUSSTAR (8C24003) of MEYS, Czech Republic. Project CLUSSTAR has received funding from the European Union’s Horizon 2020 Research and Innovation Programme under Grant Agreements No. 731473 and No. 101017733 (QuantERA). PM acknowledges a grant from the Programme Johannes Amos Comenius under the Ministry of Education, Youth and Sports of the Czech Republic reg. no. CZ.02.01.01/00/22\textunderscore 008/0004649. JF acknowledges the project 8C22002 (CVSTAR) of MEYS of the Czech Republic, which has received funding from the European Union’s Horizon 2020 Research and Innovation Programme under Grant Agreement no. 731473 and 101017733.
We thank Martin Bielak for his help in preliminary stage of the project. We thank Filip Jur\'a\v{n} for developing a 3D model shown in Fig.~2. 

\section*{Author contributions}
V.K. initiated the project, conceived the idea of cost-function-based nonlinear squeezing, and participated in data processing and numerical simulations.
J.F. and M.N. performed the experiment and data processing.
J.B. coordinated the experimental project and participated in the experiment and data processing.
P.M. introduced the concept of nonlinear squeezing and supervised the theoretical part of the project.
M.J. supervised the experimental project and participated in the experiment and interpretation.
V.K. and M.J. wrote the manuscript, and all authors were involved in revising the manuscript.
These authors contributed equally: V. Kala, J. Fadrn\'y, M. Neset.\\

\section*{Competing Interests}
All authors declare no financial or non-financial competing interests.\\

\textbf{Corresponding authors}\\
Correspondence to V. Kala and M. Je\v{z}ek.

\end{document}